\documentclass[letter]{aa} % for the letters
\usepackage{graphicx}
\usepackage{natbib}
\usepackage{epsfig}
\def\ut#1{\mathop{\vtop{\ialign{##\crcr
     $\hfil\displaystyle{#1}\hfil$\crcr\noalign
     {\kern1pt\nointerlineskip}\hbox{$\hfil\sim\hfil$}\crcr
     \noalign{\kern1pt}}}}}

\def\undersymbol#1#2{\mathop{\vtop{\ialign{##\crcr
     $\hfil\displaystyle{#2}\hfil$\crcr\noalign
     {\kern1pt\nointerlineskip}\hbox{$\hfil#1\hfil$}\crcr
     \noalign{\kern1pt}}}}}
\def\arcsec{^{\prime\prime}}

\def\degr{^0}

\begin{document}

 \title{ $XMM$ observation of 1RXS J180431.1-273932: a new M-type X-ray binary with a 494\,s-pulse period neutron star?}

      \author{A.A. Nucita
         ,
          S. Carpano, \and M. Guainazzi
          }

        \institute{XMM-Newton Science Operations Centre, ESAC, ESA, PO Box 50727, 28080 Madrid, Spain \\
         %\email{nucita@}
             }

   \offprints{A. A. Nucita}

   \date{Submitted: XXX; Accepted: XXX}

{
  \abstract
   {Low-mass X-ray binaries are peculiar binary systems composed of a compact object and a low-mass star. Recently, a new class of these systems, known as symbiotic $X$-ray binaries (with a neutron star with a M-type giant companion), has been discovered.}
  {Here, we present long-duration ${\it XMM}$ observations of the source 1RXS J180431.1-273932.}
   {Temporal and spectral analysis of the source was performed along with a search for an optical counterpart. We used a Lomb-Scargle periodogram analysis for the period search and evaluated the confidence level using Monte-Carlo simulations.}
   {The source is characterized by regular pulses so that it is most likely a neutron star. A modulation of $494.1\pm0.2$\,s (3$\sigma$ error) was found with a confidence level of $>$99\%. Evidence of variability is also present, since the data show a rate of change in the signal of $\sim -7.7\times 10^{-4}$ counts s$^{-1}$ hr$^{-1}$. A longer observation will be necessary in order to determine if the source shows any periodic behavior. The spectrum can be described by a power law with photon index $\Gamma\sim 1$ and a Gaussian line at 6.6\,keV. The X-ray flux in the 0.2--10\,keV energy band is $5.4\times 10^{-12}$ erg s$^{-1}$ cm$^{-2}$.
   The identification of an optical counterpart (possibly an M6III red-giant star with an apparent visual magnitude of $\simeq 17.6$) allows a conservative distance of $\sim 10$ kpc to be estimated. Other possibilities are also discussed.}
   {Once the distance was estimated, we got an $X$-ray luminosity of $L_X\ut<6\times 10^{34}$ erg s$^{-1}$, which is consistent with the typical $X$-ray luminosity of a symbiotic LMXB system.}
}

   \authorrunning{Nucita et al.}
   \titlerunning{$XMM$ observation of 1RXS J180431.1-273932}
   \maketitle
%
%________________________________________________________________

%%---------------------------------------------------------------

\section{Introduction}
{
Low-mass $X$-ray binaries (LMXBs) are interacting systems composed of an accreting compact object and a low-mass ($1$ M$_{\odot}$ or less) main sequence or evolved late-type star. Recently, \citet{masetti} have proposed that several faint $X$-ray sources (with typical luminosities in the range $10^{32}$-$10^{34}$ erg s$^{-1}$) in the Galaxy are wide-orbit LMXBs composed of a compact object, most likely a neutron star (NS), accreting from the wind of an M-type giant. This new kind of system is known as a {\it symbiotic} $X$-ray binary system, by analogy with symbiotic stars in which a white dwarf accretes from the wind of an M-type giant companion.

\cite{kaplan} have identified the optical counterpart of the 112 s pulsar Sct X-1=AX J183528-0737 as a red giant. However, it is unclear whether the system is a low-luminosity LMXB with a red giant or a higher luminosity binary with a red supergiant. On the other hand, the 120 s pulsar GX1+4 has been identified as an LMXB characterized by an NS accreting from an M5III red giant companion (\citealt{cha}).

Here, we report on a 100 ks {\it XMM}-Newton observation of 1RXS J180431.1-273932. The source appears to be characterized by regular pulses making it a possible NS. The optical monitor (OM) also allowed us to identify a possible optical counterpart: OGLE II DIA BULGE-SC35 4278 (a red giant star, possibly of type M6III, also found in the OGLE catalogue). We estimate the $X$-ray luminosity of the newly discovered symbiotic $X$-ray binary to be $L_X\ut<6\times 10^{34}$ erg s$^{-1}$ in the $0.2-10$ keV band. The paper is structured as follows: in Sect. 2 we describe the {\it XMM} data reduction. In Sect. 3, we give details on the spectral and timing analysis conducted on 1RXS J180431.1-273932, while in Sects. 4 and 5 we discuss the nature of the possible counterparts and draw some conclusions.}

\section{{\it XMM} Observation and data reduction}

Here, we are reporting a $100$ ks {\it XMM} observation towards the sky region around the $X$-ray (ROSAT) source 1RXS
J180431.1-273932. It was observed in October 2005 (Observation ID
30597) with both the EPIC MOS and PN cameras (\citealt{mos,pn}) operating with a thin filter
mode. The epic observation data files (ODFs) were processed using the
{\it XMM}-Science Analysis System (SAS version $7.0.0$). Using the
latest calibration constituent files currently available, we
processed the raw data with the {\it emchain} and {\it
epchain} tools to generate files with adequate event lists. After screening
with the standard criteria, as recommended by the Science
Operation Centre technical note XMM-PS-TN-43 v3.0, we rejected
any time period affected by soft proton flares. The remaining time intervals
resulted in effective exposures of $\simeq 96$ ks, $\simeq 98$ ks, and $\simeq 94$ ks for
MOS 1, MOS 2, and PN, respectively.

The coordinates of the source 1RXS J180431.1-273932, as determined by {\it edetect\_chain} tool, are
$\alpha_{\rm J2000}=18^{\rm h}04^{\rm m} 30\fs{}48$ and
$\delta_\textrm{J2000}=-27^\circ 39' 32\farcs 76$.
A systematic shift (X-ray--optical) of $-2.52\arcsec$ ($1\sigma$) and $-3.09\arcsec$ ($1\sigma$) exists in right ascension for MOS 1 and MOS 2 and for $-1.19\arcsec$ and $0.41\arcsec$ in declination, respectively.
We assume in the following that the error associated to the source coordinates, as determined
by the $X$-ray observation, is at least $\sim 2\arcsec$ on both celestial coordinates.

\section{Spectral and timing analysis of 1RXS J180431.1-273932}

The source spectra were extracted in a circular region centered on the nominal position of the target in the
three EPIC cameras, while the background spectra were accumulated in annuli on the same coordinates.
The resulting spectra were rebinned to have at least
25 as the minimum number of counts per energy bin. The EPIC PN image has to be taken with caution since the source is on a chip gap, hereby reducing the net number of collected $X$-ray photons.

\begin{figure}[htbp]
\vspace{7.0cm} \includegraphics{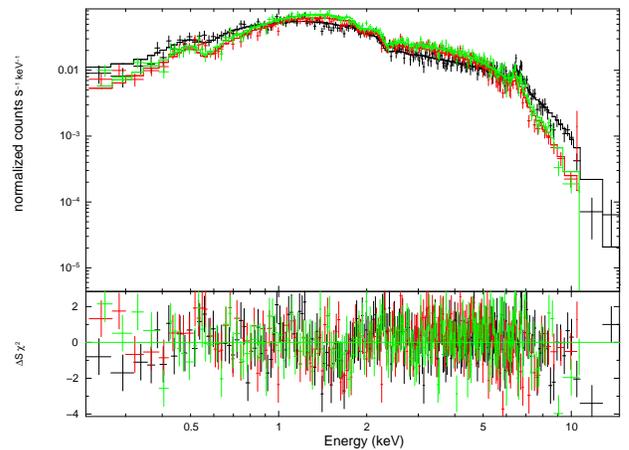}
\caption{The simultaneous fit to the MOS 1, MOS 2, and PN data with a model
constituted of an absorbed power law to which an emission line was added.}
\label{fig2}
\end{figure}

The spectra were simultaneously fitted with XSPEC (version 12.0.0). In Fig. \ref{fig2},
we show the MOS 1, MOS 2, and PN spectra for 1RXS J180431.1-273932 and the respective fits. Note the clear evidence of an iron emission line in the spectra. The best-fitting model was an absorbed power law to which a Gaussian line is added. We left all the parameters free, which yielded $\chi^2/\nu$=1.19 for $\nu=611$ degrees of freedom.
The results of spectral fitting are shown in Table \ref{table2}. Uncertainties are given at a $90\%$ confidence level.
There, $N_H$ is the equivalent column density of neutral hydrogen, $\Gamma$ the photon index, $E_L$ the energy of the line, $\sigma_L$ its width, and $N_L$ and $N_{\Gamma}$ the normalization of the Gaussian line and the power-law
component, respectively. The corresponding $0.2-10$ keV flux is shown in the last column.
Both the location and width of this line ($6.59$ keV and $0.23$ keV) are indications that we are dealing
with a blending of multiple lines that, unfortunately, cannot be resolved.
{We also tried to fit the data with different models but without obtaining a substantial decrease in the reduced $\chi^2$ value. In particular, by requiring that the column density value matches what is obtained by the $N_H$ calculator\footnote{http://heasarc.nasa.gov/cgi-bin/Tools/w3nh/w3nh.pl} in the direction of the target, i.e. $N_H\simeq 3.4\times 10^{-21}$ cm$^{-2}$, a three-component model (thermal bremsstrahlung, power law, and Gaussian line) is needed. In this case, the resulting best-fitting parameters are $KT=(0.26\pm0.02)$ keV and $\Gamma=1.15\pm0.02$, while the remaining parameters are practically the same as in Table 1. In this case, one has $\chi^2/\nu$=1.18 for $\nu=609$ degrees of freedom}

\begin{figure}[htbp]
\vspace{5.8cm} \includegraphics{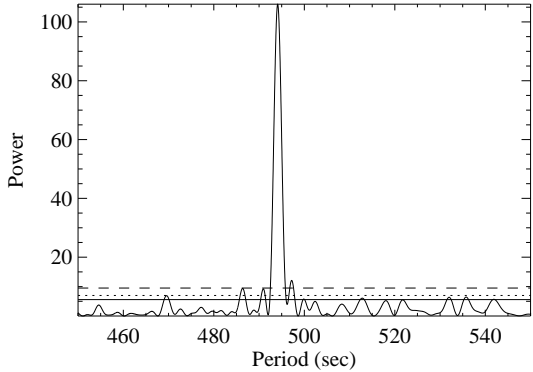}
\caption{Search for periodicities in the {\it XMM}-Newton light curve of 1RXS~J180431.1-273932  using a
 Lomb-Scargle periodogram analysis. The full, dotted, and dashed lines represent the 68\%, 90\%, and
 99\% confidence level, respectively.}
\label{fig:period}
\end{figure}

\begin{figure}[htbp]
\vspace{5.0cm} \includegraphics{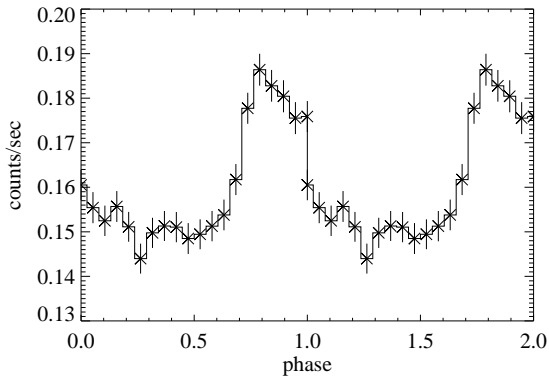}
\caption{The 0.3--8.0\,kev light curve folded at 494.1\,s using 20 bins.
 Phase zero is associated to the
 beginning of the {\it XMM}-Newton observation.}
\label{fig:fold}
\end{figure}

The source 1RXS J180431.1-273932 has been observed in the past by the ROSAT satellite with an exposure
time of $255$ s. The main observational characteristics are thus available in the ROSAT All-Sky Survey Catalogue
where the source shows an $X$-ray count number of $6.14\times 10^{-2}$ counts s$^{-1}$. By using PIMMS, for a power-law model with photon index $\Gamma=1$ and column density  $N_H \simeq 10^{21}$ cm$^{-2}$, one gets a flux of $5.618\times 10^{-12}$ erg cm$^{-2}$s$^{-1}$ (in the 0.2-10 keV band) corresponding to an unabsorbed flux of $6.159\times 10^{-12}$ erg cm$^{-2}$s$^{-1}$, consistent with what was derived by the XMM observation.
\begin{table}
\begin{center}
\caption{Results of the spectral fits for 1RXS J180431.1-273932,
using an absorbed power law model and a Gaussian line (see text for details).}\label{table2}
\begin{tabular}{llllll}
\hline
\multicolumn{6}{c}{1RXS J180431.1-273932: best fit values}\\
\hline
 $N_H$$\times10^{21}$ ({\rm cm$^{-2}$})            & $1.60_{-0.10}^{+0.10}$ &\\
 $\Gamma$                                          & $1.07_{-0.02}^{+0.02}$ &\\
 $N_{\Gamma}$$\times 10^{-4}$                      & $3.41_{-0.11}^{+0.10}$ &\\
 $E_L$ (keV)                                       & $6.59_{-0.05}^{+0.05}$ &\\
 $\sigma _L$ (keV)                                 & $0.23_{-0.04}^{+0.04}$ &\\
 $N_{L}$$\times 10^{-5}$                           & $1.64_{-0.36}^{+0.39}$ &\\
 $F_{0.2-10 {\rm keV}}$$\times 10^{-12}$({\rm cgs})& $5.41_{-0.30}^{+0.20}$ &\\
\hline
\end{tabular}
\end{center}
\end{table}
We analyzed the light curve of 1RXS~J180431.1-273932 and searched for a period signal between 5 sec and 10 \,h.
The long-term light curve is flat (but see next section), and we found a periodic signal at 494.1\,s using a
Lomb-Scargle periodogram (\citealt{Lomb1976,Scargle1982}). By means
of Monte Carlo simulations, we evaluated the confidence level by assuming a null hypothesis of white noise, and the
results are plotted in Fig.~\ref{fig:period}, with the 68\%, 90\%, and 99\% confidence levels.
We found that the 494.1\,s period is significant at a confidence level $>99\%$.
To estimate the error, we fitted the light curve with a sine function using the IDL task \texttt{curvefit},
keeping the trial periods fixed. The method has been described in more detail in
\cite{Carpano2007}. The 3$\sigma$ error of the 494.1\,s pulse period is 0.2\,s.
The XMM light curve folded at 494.1\,s is shown in  Fig.~\ref{fig:fold}.

\begin{figure}[htbp]
\vspace{5.3cm} \includegraphics{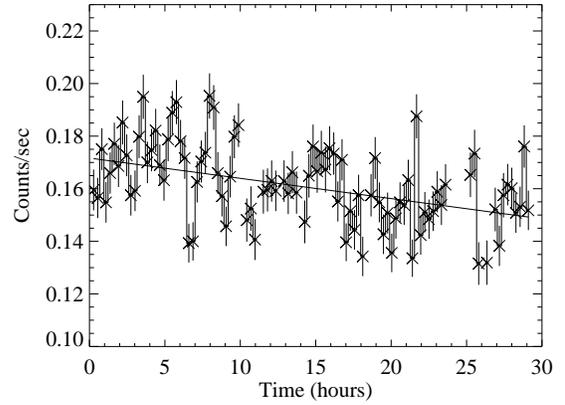}
\caption{The light curve of the 100 ks observation as binned at twice the pulse period of $494.1$ s. The observed signal decreases within the observation duration by the amount of $-(7.7\pm0.9)\times 10^{-4}$ counts s$^{-1}$ hr$^{-1}$ (see text for more details).}
\label{fig6}
\end{figure}

{The light curve of the 100 ks observation was then binned at twice the pulse period (to reduce the noise) to search for variabilities in the $X$-ray source. As one can note from Fig. \ref{fig6}, the observed signal decreases on the observation time scale from the value of $\sim 0.17$ count s$^{-1}$ to $\sim 0.15$ count s$^{-1}$. By a linear fit to the binned data, one obtains a rate of change in the observed signal of
$-(7.7\pm0.9)\times 10^{-4}$ counts s$^{-1}$ hr$^{-1}$. Note also that the light curve shows irregular variability ($>20\%$) on time scales of hours. Of course, a longer observation is necessary to clarify whether the change in luminosity has a periodic behavior.}

\section{The nature of 1RXS J180431.1-273932: a possible optical counterpart}

To get more information about 1RXS J180431.1-273932 and to address some
hypotheses on its nature, we searched for possible optical counterparts (within a few
arcseconds from the nominal position of the X-ray source) in available catalogues. Among
over 200,000 Galactic bulge variable stars contained in the public domain OGLE catalogue (\citealt{wep03}),
we found that a source exists within $\simeq 4.8$ arcseconds of 1RXS J180431.1-273932 and
has been identified as a variable red giant and labeled as
OGLE II DIA BULGE-SC35 4278. The object is found in a large catalogue that was compiled earlier of stars that trace
the galactic bar. In fact, the
average magnitudes for these stars correlate well with the Galactic longitude and vary from
$I_{k,0}=11.82$ for $l=+8^{\degr}$ to $I_{k,0}=12.07$ for $l=-5^{\degr}$, clearly indicating that they are located in
the Galactic bar inclined to the line of sight. {These stars form three well-separated groups: luminous red giants, less luminous red giants located near the bulge and relatively blue Galactic disc stars (with only a small fraction of the observed stars identified as foreground blue objects).}

{In Table \ref{table4}, the main photometric properties of the optical counterpart
are reported as given in the OGLE catalogue. OGLE II DIA BUL-SC35-4278 is recognized as a pulsating red giant with a main period of $20.26$ days and an amplitude of $0.02$ mags in the I band. In this case, it falls within the range given for A-type variables in Eq. 1a of the above-mentioned paper and are believed to be bright giants.

Using the reddening maps towards the Galactic center provided by \citet{sumi}, \citet{wep03} find that the extinctions in the $V$ and $I$ bands are $A_V\simeq 1.915$ and $A_I\simeq 0.940$, respectively. The extinction terms in the other bands were determined using the standard relations in Table 6 of \citet{sfd1998}, i.e. $A_J/A_V = 0.276$, $A_H/A_V = 0.176$, and $A_K/A_V = 0.112$, thus giving $A_J\simeq 0.528$, $A_H\simeq 0.337$, and $A_K\simeq 0.214$, respectively. From Table \ref{table4}, the dereddened colors of the source can be obtained by applying the magnitude correction defined above. In particular, one gets $(V-K)=6.82$, $(J-H)=0.863$, and $(H-K)=0.305$ (typical of a red star rather than a reddened blue star) so that the source spectral type can be inferred to be close to an M6III giant star using well known tables of color indices (\citealt{allen}). Of course, we are assuming here that the source is a giant star due to its position in the HR diagram for bulge stars.

On the other hand, the star might be a very reddened, intrinsically blue object; but this possibility is unlikely since we do not expect to see blue stars in the bulge, and any observed early-type star should be the foreground to the bulge. In addition, an early type star needs to be more reddened than a red star to be observed with the given colors.
Of course, classification of the detected optical counterpart as an M6III red giant or as a strongly reddened blue star will be clear once the characteristic of its spectrum is available in the near future. We are in fact planning to request observational time in order to get a spectrum of OGLE II DIA BULGE-SC35 4278 and define its nature.

We searched for optical counterparts in the OM and found that,
within $3.4\arcsec$ of the $X$-ray source position, an object with instrumental $v$ magnitude of $\simeq 17.2$ exists.
In Table \ref{table3}, we give the coordinates of the OM identified source and the coordinates of the same source as
found in the OGLE catalogue. Note that the position shift between the OM source position and the catalogue
target coordinates is due to the strong aberration (up to $4\arcsec$, \citealt{cal-tn-0019-3-2}) affecting the OM
at the field of view border.

Assuming that a star of spectral type M6III has an absolute magnitude $M_V\simeq -0.2$ (\citealt{lang}) and by using the extinction corrected magnitude $m_V\simeq 15.753$, we estimated the source distance to be $\sim 15$ kpc. Note that \citet{wep03} hypothesize that most of the sources in their sample are in the Galactic bulge, as confirmed by the observation  where they trace the Galactic bar. However, according to maps in \citet{sumi}, the typical visual extinction in this field can vary from $1.4$ to $2.3$. In addition, we can say that, if the extinction is underestimated by a few tenths of a magnitude, then the distance may be overestimated by a few kpc \footnote{The measured column density value (derived for a power-law model, see Table 1) is lower than what is expected ($\sim 4\times 10^{21}$ cm$^{-2}$) in a direction with extinction $A_V\simeq 1.9$. One intriguing possibility is that the source is a foreground cataclysmic variable created by a white dwarf interacting with a very late companion that mimics a bulge star. In this case, the periodicity of $494$ s (observed in the $X$-ray signal) would be the spin period of the white dwarf.
However, by using the empirical relation between the extinction and column density values (\citealt{av_nh}), one gets $A_V\simeq 0.8$ for the measured $N_H$, and the dereddened color indices seem to be incompatible with a main sequence, late type star. Again, only an optical spectrum will give a definitive answer.}. Therefore, to be conservative we assume $d\simeq 10$ kpc so that, with the $0.2-10$ keV flux determined in the previous section, one gets a source $X$-ray luminosity of $L_X=F_{0.2-10 {\rm keV}} 4 \pi d^2 \ut< 6\times 10^{34}$ erg s$^{-1}$, which is consistent with the typical $X$-ray luminosity of an LMXB composed of an M-type giant interacting with an NS.

\begin{table}
\begin{center}
\caption{Optical counterpart position as determined by the OM data and as found in the available catalogues.}\label{table3}
\begin{tabular}{lll}
\hline & RA(J2000)   & DEC(J2000)\\
\hline
{\rm OGLE cat.}  &$18:04:30.14$& $-27:39:34.20$\\
{\rm OM}         &$18:04:30.48$& $-27:39:34.19$\\
\hline
\end{tabular}
\end{center}
\end{table}

\begin{table}
\begin{center}
\caption{The observed source magnitudes and color indices in standard bands as in the OGLE catalogue.}\label{table4}
\begin{tabular}{lllllll}
\hline
   $m_V$& $(m_V-m_I)$& $m_I$& $m_J$ & $m_H$ & $m_K$\\
\hline
   17.668& 4.397 & 13.263 & 10.627 & 9.573 & 9.145 \\
\hline
\end{tabular}
\end{center}
\end{table}

\begin{figure}[htbp]
\vspace{5.2cm} \includegraphics{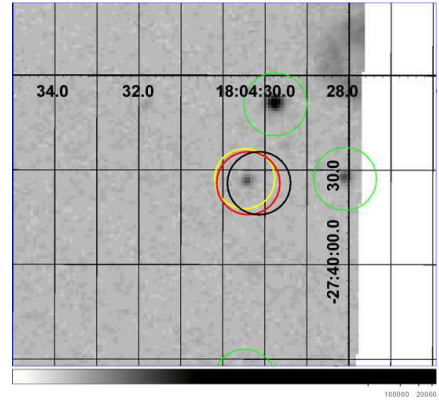}
\caption{The OM image centered on the position of the $X$-ray source (yellow circle).
The red ring is centered on the OM counterpart, while the black
one is at the optical counterpart coordinates given in the OGLE catalogue. The radius of each ring is $10\arcsec$. Other close sources are
identified by green circles.}
\label{fig5}
\end{figure}

}

\section{Discussion}

{LMXBs are thought to be binary systems containing a compact object interacting with a low-mass star. Interestingly, a new class of LMXBs has recently been discovered: a neutron star possibly in a wide orbit around
an M-type giant. This new $X$-ray source is currently known as a symbiotic $X$-ray binary.

In this paper, we have reported on a 100 ks {\it XMM}-Newton observation of 1RXS J180431.1-273932. The source appears to be characterized by regular pulses, with a period of
$(494.1\pm 0.2)$\,s ($3\sigma$ error), so that it is probably an NS.

The OM identified a possible optical counterpart (a giant star), also found in the OGLE catalogue
where it is labeled as OGLE II DIA BULGE-SC35 427. The source appears to be a giant star (an M6III star with magnitude in the optical band of $\simeq 17.6$) so that, if it is interacting with the identified $X$-ray object, it is likely that a new symbiotic $X$-ray system has been discovered in our Galaxy. We estimated the distance of the binary system to be $d\sim 10$ kpc. Hence, by using the observed $XMM$ flux of $F_{0.2-10 {\rm keV}}=5.41\times 10^{-12}$ erg s$^{-1}$ cm$^{-2}$, we get an $X$-ray luminosity of $L_X\ut<6\times 10^{34}$ erg s$^{-1}$, which is consistent with the typical $X$-ray luminosity of a symbiotic LMXB system.

}

\begin{acknowledgements}
This paper is based on observations from XMM-Newton, an
ESA science mission with instruments and contributions directly funded by ESA
Member States and NASA. We are enormously grateful to the anonymous referee for the corrections
that greatly improved the work. One of us (AAN) is grateful to Dr. F. De Paolis, Dr. J. Wray, and Dr. L. Eyer for the precious suggestions.
\end{acknowledgements}

%----------------------------------------------------------------------

\end{document}